\documentclass[preprint,12pt]{elsarticle}
\usepackage{epsfig}
\usepackage{graphicx}
\usepackage{amsmath,amsfonts,amssymb}
\newcommand{\be}{\begin{equation}}
\newcommand{\ee}{\end{equation}}
\newcommand{\bea}{\begin{eqnarray}}
\newcommand{\eea}{\end{eqnarray}}
\newcommand{\bx}{{\bf x}}
\newcommand{\by}{{\bf y}}
\newcommand{\bk}{{\bf k}}

\newcommand{\bn}{{\bf n}}

\journal{Physica A}

\begin{document}

\begin{frontmatter}



\title{Peaks in the CMBR power spectrum. I. Mathematical analysis of the
associated real space features}


\author{M. L\'opez-Corredoira$^{1,2}$ \& A. Gabrielli$^{3,4,5}$}

\address{$^1$ Instituto de Astrof\'\i sica de Canarias,
E-38200 La Laguna, Tenerife, Spain\\
$^2$ Departamento de Astrof\'\i sica, Universidad de La Laguna,
E-38206 La Laguna, Tenerife, Spain\\
$^3$ Istituto dei Sistemi Complessi CNR, Via dei Taurini 19, 00185 Rome,
Italy\\ 
$^4$ IMT Institute of Advanced Studies, Piazza S. Ponziano 6, 55100 Lucca, Italy \\
$^5$ London Institute of Mathematical Science, 35a South St,
London W1K 2XF, UK}

\begin{abstract}
The purpose of our study is to understand  the mathematical origin in real space of modulated and damped 
sinusoidal peaks observed in cosmic microwave background radiation anisotropies. We use the theory of the Fourier 
transform to connect localized features of the two-point correlation function in real space to oscillations 
in the power spectrum. We also illustrate analytically and by means of Monte Carlo simulations the angular
 correlation function for  distributions of filled disks with fixed or variable radii capable of generating 
 oscillations in the power spectrum. While the power spectrum shows repeated information in the form of  
 multiple peaks and oscillations, the angular correlation function offers a more compact presentation that condenses 
 all the information of the multiple peaks into a localized real space feature. We have seen that oscillations 
 in the power spectrum arise when there is a discontinuity in a given derivative of the angular correlation 
 function at a given angular distance. These kinds of discontinuities do not need to be abrupt in an infinitesimal
  range of angular distances but  may also be smooth, and  can be generated by simply distributing excesses 
  of antenna temperature in filled disks of fixed or variable radii on the sky, provided that there is a 
  non-null minimum radius and/or the maximum radius is constrained.
\end{abstract}


\end{frontmatter}

\section{Introduction}

We present a series of two papers with the motivation of
understanding the CMBR (''Cosmic Microwave Background Radiation'') 
anisotropies\cite{Whi94} without  direct comparison with a physical model, in order to see whether other 
conditions different from the standard model of cosmology may generate something similar of what is observed. 
In this first part, we will focus qualitatively on the mathematical elements necessary to produce peaks in the 
power spectrum of the CMBR, and in the second part \cite{Lop12} we will discuss its physical generic 
interpretation and  evaluate
quantitatively the number of free parameters a generic theory needs in order to fit it.
This line of research is at present almost totally abandoned, the available literature 
merely discussing how the peaks can be generated in terms of the standard cosmology. We could consider our theme as 
an inversion problem: rather than deriving the predictions from a model to fit the data, we wish to ascertain from 
the data some general characteristics of a generic model.
Our goal is to open analysis to wider mathematico--physical scenarios, in order to see whether something 
different from standard cosmology interpretation can reproduce it.

The analysis of  CMB maps is usually performed in terms of 
spherical harmonic decomposition and by computing the angular power
spectrum of CMB anisotropies \cite{Tri05,Pei10,Lar11,Das11}. The standard
cosmological model provides a physical model able to fit accurately
the power spectrum of CMB anisotropies. It is well known that this
exhibits a series of peaks, or an almost periodic oscillation, as 
 predicted by Peebles \& Yu \cite{Pee70}.

In the ideal case of full-sky coverage, the
two-point angular correlation function provides a complementary means
of analyzing CMB observations instead of the angular power spectrum and, in principle,
contains the same information as the angular power
spectrum. However, as we will discuss in what follows, the angular
correlation function allows an easier understanding of the anisotropy
structures and it may serve as a complementary means of spherical
harmonic analysis.
Few authors have considered the direct determination of the
anisotropies correlation function directly in angular space 
\cite{Smo92,Hin96,Kas01,Her04,Cop06,Cop07,Cop09},
(see the review in ref. \cite{Cop10}), and many of them 
have focused on the large scale
behavior of the angular correlation function, where the effect of
cosmic variance is certainly important. 
In particular, the two-point correlation
function almost vanishes on scales greater than about 60 degrees,
contrary to what the standard CDM theory predicts, and  is in agreement
with the same finding obtained from COBE data about a decade earlier\cite{Smo92,Hin96}. 
It was pointed out\cite{Cop10} that the
striking feature of the two-point angular correlation function is the
lack of large angle correlations, unexpected in inflationary models.
On the other hand it is also
interesting to consider, both theoretically and observationally, the
behavior of the correlation function on small angular scales \cite{Bon87,Bas02}. Indeed,
the whole structure of the peaks in the angular power spectrum
corresponds in direct angular space to a localized feature of the
correlation function which is  analogous for the CMB field to the
baryon acoustic oscillation (BAO) scale characterizing the matter
correlation function \cite{Eis05,Kaz10} (see criticism on its
detection in ref. \cite{Syl09}).  
The direct identification of such a scale would thus represent a complementary
test of the standard model.

In this paper, we illustrate some simple mathematical properties of
the anisotropy structures  associated with both the
oscillating peaks in the angular power spectrum and with the characteristic 
scale in the anisotropy angular distribution. In particular, we
provide a simple explanation of the peak sequence, showing that it is simply
associated with the shape of  the structures of the  anisotropies. While this
discussion is useful to illustrate the statistical meaning of the
properties of the $\Lambda $-CDM angular correlation function, it also provides
a simple general model
to generate such type of correlations.

Our paper is structured as follows: In \S \ref{.oscill}, we discuss
the mathematical origin of the oscillating peaks in the power
spectrum, and of the corresponding scales in the angular
distribution of anisotropies. In \S \ref{.examples}, we illustrate the
situation by introducing a few simple toy models able to show similar
correlation properties as those predicted by standard cosmological
models. Finally, in \S \ref{.concl} we summarize the results.

\section{The origin of oscillations in the power spectrum and its correspondence in the
correlation function at small angles}
\label{.oscill}

\subsection{Power spectrum and self-correlation}

The fluctuation field on the sky, $\frac{\delta T}{T}(\hat\bn)$, where $\hat\bn$ is the general point on the 
unit sphere (or direction in  $3-$dim space), can be decomposed into
spherical harmonics in the sphere:
\begin{equation}
\frac{\delta T}{T}(\hat\bn)=\sum _{\ell=0}^\infty \sum
_{m=-\ell}^{+\ell}a_{\ell m}Y_{\ell m}(\hat\bn) \;.
\end{equation}
Assuming statistical isotropy, the  
angular power spectrum $C_\ell$ is 
defined as
\begin{equation}
C_\ell \equiv \langle |a_{\ell m}|^2\rangle=\frac{1}{2\ell +1}
\sum _{m=-\ell }^{+\ell }|a_{\ell m}|^2 \;, 
\end{equation}
whereas the angular two-point correlation function is defined as [\cite{Pad93},\cite{Gab05}(\S 6.6.1)]:
\be
C(\theta )\equiv \left \langle \frac{\delta T}{T}(\hat\bn)\frac{\delta T}{T}(\hat\bn')\right \rangle 
\ee
where $\theta$ is the separation angle between the two directions $\hat\bn$ and $\hat\bn'$.  
The two-point correlation function can be expanded in 
terms of Legendre polynomials ($P_\ell $):
\be
C(\theta ) = \frac{1}{4\pi }\sum _{\ell =0}^\infty (2\ell+1)C_\ell
P_\ell [\cos (\theta )] \;.
\label{ctheta}
\ee

By inverting Eq. (\ref{ctheta}) we find\cite{Cop10}
\begin{equation}
C_\ell =2\pi \int _0^\pi C(\theta )P_\ell [\cos (\theta )]\sin (\theta )d\theta .
\label{cl}
\end{equation}

Alternatively, always assuming statistical isotropy, we can express the power spectrum
in terms of the conjugate frequency $k$ in the Fourier transform:

\be
P(k)=\tilde C(k) \equiv FT[C(\theta )]= 
\int d\vec{\theta } C(\theta )e^{-i\vec{k}\vec{\theta }}
,
\ee
where the vectors $\vec{\theta }$ and $\vec{k}$ are  two-dimensional vectors whose 
moduli are $\theta $ and $k$ respectively, and  

\begin{equation}
C(\theta )=FT^{-1}[P(k)] \;. 
\end{equation}
For small enough angles $\theta \ll 1$ radians, 
$P_\ell [\cos (\theta )]\approx J_0[(\ell  +1/2)\theta ]$ \cite{Pee80}(Eq. 46.39), 
where $J_0$ is the Bessel function of order zero, and 
$P(k)\approx 2\pi \int _0^\pi C(\theta )$ $J_0(k\theta )\sin \theta d\theta$ \cite{Pee80}(Eq. 46.43),
so the spherical harmonic decomposition coincides with the two-dimensional Fourier transform 
with a frequency $k= (\ell +1/2)$. From now on, we will not distinguish between $C_\ell $ and $P(k)$.

\subsection{Relationship between oscillations in the power spectrum and abrupt transitions
in the self-correlation function}
\label{.mat1}

In order to clarify the relation between the sequence of
oscillating peaks in the CMB angular power spectrum and the angular
space two-point correlation function, let us introduce a simple
model. We first distribute points centers uniformly in space
and then distribute continuous mass profiles around
these points, which hereafter we denominate as centers.  
Since we are interested in the angular correlation function $C(\theta)$ at
$\theta\ll 1$ radians, we can approximate the angular space as Euclidean and use Fourier 
transform analysis to illustrate mathematically the link between sinusoidal
oscillations in the power spectrum and strongly localized features in the
two-point correlation function.  

In general
in a $d-$dimensional Euclidean space the model can be formulated as
follows.
Let us start with a stochastic spatial distribution of centers 
 with microscopic number density 
\be
n(\bx)=\sum_{i}\delta(\bx-\bx_i)\,,
\label{density} 
\ee 
where the sum runs over all center positions  $\bx_i$.  
Let us assume that this distribution is statistically homogeneous in space, which implies: 
(i) the average number density is a positive constant $n_0>0$, and (ii) the two-point correlation 
function (TPCF) depends only on the separation vector between these two points.  
In this case, the TPCF can be defined as \cite{Gab05} 
\be  
\xi(\bx)=\frac{\left<n(\bx_0)n(\bx_0+\bx)\right>-n_0^2}{n_0^2}= 
\frac{\delta(\bx)}{n_0}+h(\bx)\,, 
\label{n-cov} 
\ee where $\left<\cdot\right>$ represents the ensemble average, or the
infinite volume average in the case of ergodicity.  We assume that
$h(\bx)$ (the covariance or off-diagonal correlation function) is a
sufficiently regular function with no discontinuity in any of its
derivatives for $x>0$ and decays sufficiently rapidly at large separations
$\bx$ (e.g., one can suppose an analytic function in all 
space $\bx$ as an exponential or power law decay).

Due to the assumed statistical and spatial homogeneity, the power spectrum $S(\bk)$ of the
stochastic distribution of centers satisfies  
\be 
S(\bk)\equiv\lim_{V\to\infty}\frac{\left<|\delta 
\tilde n(\bk;V)|^2\right>}{n_0V}= 1+n_0\tilde h(\bk)\,, 
\label{n-ps} 
\ee 
where  
\bea 
&&\delta \tilde n(\bk;V)=\int_Vd^dx[n(\bx)-n_0]e^{-i\bk\cdot\bx}\\ 
&&\tilde h(\bk)=FT[h(\bx)]=\int_{\mathbb{R}^d} d^dx\,h(\bx)e^{-i\bk\cdot\bx}\,. 
\label{FTs} 
\eea  
Note that, since $\tilde h(\bk)$ vanishes at large $k$, $S(\bk)$ in the same
limit converge to $1$.
As shown more rigorously below, since $h(\bx)$ is thought to have all 
continuous derivatives at $x>0$, $S(\bk)$ is expected not to 
present damped sinusoidal oscillations.

Let us now replace each center at $\bx_i$ with a 
continuous density profile described by the function $f(\bx-\bx_i)\ge 0$, such that 
$f(\bx\to\infty)\to 0$ sufficiently fast so that $\int_{\mathbb{R}^d} d^dx\,f(\bx)$ 
is equal to a finite positive constant which we can take to be $1$.  
The field can now be written as 
\be  
T(\bx)=\sum_i f(\bx-\bx_i)\,.
\label{m-density} 
\ee 
The mean value is clearly  
$\left<T(\bx)\right>=n_0\int d^dx\,f(\bx)$.  

Since we can write 
\begin{equation}
T(\bx)=\int d^dy f(\bx-\by)n(\by)\,,
\end{equation} 
it is simple to show that the power spectrum $P(\bk)$ of the 
field $T(\bx)$ is simply given by 
\be 
P(\bk)=|\tilde f(\bk)|^2S(\bk)\,, 
\label{m-ps} 
\ee 
where $\tilde f(\bk)=FT[f(\bx)]$. Note that, as $f(\bx)$ is
assumed to be integrable, $\tilde f(\bk)$ vanishes at large $k$ and therefore also
$P(\bk)$ vanishes at large $k$. At the same time, the TPCF of the
field $T(\bx)$ is a continuous function, as it has to be for any proper
continuous field\cite{Gab05}.

Since we have assumed that $S(\bk)$ is not characterized by oscillating peaks, if we
want $P(\bk)$ to show them, we need these to be associated with $\tilde f(\bk)$. 
Thus, the problem is reduced to finding which features of  $f(\bx)$ generate oscillating peaks 
in $P(\bk)$. We show below that if $f(\bx)$ presents a finite discontinuity
in the function or in one of its derivatives at a position $\bx_0\ne 0$,
then $P(\bk)$ will show slowly damped sinusoidal oscillations and 
wave-vector peaks (in Fourier space) $\bx_0$.

A simple example is given by the following function $f(\bx)$ in $d$ dimensions, usually called 
{\em spherical box function}
\[f(\bx)=\left\{
\begin{array}{ll}
1/\|S_d(R)\| & \mbox{for } \bx\in S_d(R)\\
0 & \mbox{otherwise}
\end{array}
\right.
\]
where $S_d(R)$ is the $d-$dimensional sphere of radius $R$ centered at the origin and 
$\|S_d(R)\|$ its volume. In this case, $\tilde f(\bk)$
is characterized by slowly damped sinusoidal oscillations:
\begin{equation} 
\tilde f(k)=\left \{ 
\begin{array}{ll}
\frac{\sin (kR)}{kR},& \mbox{$d=1$} \\
\frac{2}{kR}J_1(kR),& \mbox{$d=2$} \\
3\frac{\sin(kR)-kR\cos(kR)}{(kR)^3},& \mbox{$d=3$}
\end{array}
\right \} \;, 
\end{equation}
where $J_1$ is the Bessel function of the first kind.  If, for example, we
have a Poissonian distribution of centers, i.e., one with $S(\bk)=1$, we
have simply $P(\bk)=| \tilde f(k)|^2$ which shows slowly damped
sinusoidally oscillating peaks.
A generalization of this development for a variable radius $R$ is given in
the following subsection.

\subsection{Mathematical description of the origin of oscillations in the varying radius case}
\label{.mat2}

We can generalize the mathematical description of \S \ref{.mat1} for variable 
radius balls by considering the following field
\be
T(\bx)=\sum_i f(\bx-\bx_i;R_i)\,,
\label{rho2}
\ee
where $\{\bx_i\}$ are the positions of the centers and
\[f(\bx;R_i)=A(R_i)g(x)\theta(R_i-x)\,,\]
where the radius $R_i$ changes from center to center, $g(x)$ is an arbitrary
universal non-negative function and $A(R_i)$ is the probability density function.

In order to find the power spectrum $P(\bk)$ we should calculate the following
double average
\[P(\bk)=\frac{\left<\overline{|\delta\tilde T(\bk,V)|^2}
\right>}{T_0V}\,,\]
where $\delta\tilde T(\bk,V)=FT_V[T(\bx)-T_0]$, $\left<\cdot\right>$ is
the ensemble average over the positions of centers and $\overline{(\cdot)}$ is
the average over the distributions of radii. The two averages clearly commute.
The first step is to write
\[\tilde T(\bk,V)=\int_Vd^dx T(\bx)e^{-i\bk\cdot\bx}=\sum_ie^{-i\bk\cdot\bx_i}
\tilde f(\bk;R_i)\,.\]
We can then write
\[|\tilde T(\bk,V)|^2=\sum_{i,j}\tilde f(\bk,R_i)\tilde f(\bk,R_j)
e^{-i\bk\cdot(\bx_i-\bx_j)}\,.\] It is simple to show that the
diagonal part of this double sum is exactly balanced by the
subtraction of the mean density from $T(\bx)$ in the actual
definition of the PS, for this reason only the non-diagonal part
contributes to the final PS.
 We have therefore to calculate the following average
\be
\overline{|\tilde T(\bk,V)|^2}=\left[\overline{\tilde f(\bk;R)}\right]^2
\sum_{i,j}'e^{-i\bk\cdot(\bx_i-\bx_j)}\,,
\label{PS-vary}
\ee
where the prime in the double sum means the condition $i\ne j$, and
\[\overline{\tilde f(\bk;R)}=\int d^dx\, e^{-i\bk\cdot\bx}
g(x)\int_{-\infty}^{+\infty} dR\,A(R)\theta(R-x)\,.\]
By taking the second average $\left<\cdot\right>$ of Eq.~(\ref{PS-vary}), 
finally we get
\[P(\bk)=\left[\overline{\tilde f(\bk;R)}\right]^2S(\bk)\,.\]
Thus the problem is reduced to calculating $\overline{\tilde f(\bk;R)}$.

In other words, the power spectrum is the superposition of the power
spectra for each value of $R_i$.

\subsection{General relation between singular points in a function and damped oscillations in its FT}

Let us now give the general argument relating the discontinuity of the function 
$f(\bx)$, or of one of its derivatives at some point $\bx_0\ne 0$ to
the presence of slowly (i.e., power law) damped oscillations (and peaks) in
$P(\bk)$. With the aim of simplicity let us develop the argument in $d=1$ 
dimensions. Let us assume that the function $f(x)$ is analytic everywhere with the exception 
of the point $x_0$ where its $n^{th}$ derivative 
$f^{(n)}(x)$ presents a finite discontinuity: $f^{(n)}(x_0^+)-f^{(n)}(x_0^-)=A$.
Moreover, let us assume that $f(x)$ vanishes sufficiently rapidly at
large $|x|$. Let us study the features of the FT of $f(x)$ induced by
such discontinuities. The FT is defined by
\begin{equation}
\tilde f(k)=\int_{-\infty}^{+\infty}dx\,e^{-ikx}f(x)\,.
\end{equation}
By integrating it by parts $n+1$ times, we get
\bea
\label{delta}
&&\tilde
f(k)=\frac{1}{(ik)^{n+1}}\int_{-\infty}^{+\infty}dx\,e^{-ikx}f^{(n+1)}(x)\\ 
&&=\frac{1}{(ik)^{n+1}}\lim_{\epsilon,\epsilon'\to 0}\left[\int_{-\infty}^{x_0-\epsilon}
dx\,e^{-ikx}f^{(n+1)}(x)+ \right.\nonumber\\
&&\left.\int_{x_0+\epsilon'}^{+\infty}dx\,e^{-ikx}f^{(n+1)}(x)
+A\int_{x_0-\epsilon}^{x_0+\epsilon'}dx\,e^{-ikx}\delta(x-x_0)\right]\nonumber\\
&&=\frac{1}{(ik)^{n+1}}\int_{-\infty}^{+\infty}
dx\,e^{-ikx}g(x)+\frac{Ae^{-ikx_0}}{(ik)^{n+1}}\,,\nonumber
\eea
where
\[g(x)=f^{(n+1)}(x)-A\delta(x-x_0)\]
is an analytic function everywhere but in $x_0$
where at most it can have a finite discontinuity. In this last case
this discontinuity will lead to a further, but subdominant
(i.e., decreasing faster with $k$) contribution to the slowly damped
oscillations given by the last term in Eq.~(\ref{delta}).  It is important to note that the 
power of decay of damped oscillations is directly related to the order of the discontinuous derivative.

This argument can be extended to higher dimensions. For a statistically isotropic 
distribution in $d$
dimensions in which both $f(\bx)$ and $\xi(\bx)$ depend only on
$x=|\bx|$, and consequently $\tilde f(\bk)$, $S(\bk)$ and $P(\bk)$
depend only on $k=|\bk|$. In this case, when $f(\bx)\equiv f(x)$ shows
a discontinuity in only one of the derivatives in $x$, 
the calculation can be reduced to a one-dimensional one very similar to the 
one-dimensional case treated above.

The terms $\exp(-ikx_0)$ in Eq. (\ref{delta}) give the oscillations.
Thus oscillating peaks in the power spectrum occur when the
correlation function has an abrupt change of functionality at some
$\theta _0>0$; that is, when there is a characteristic angular
distance which separates two regimes in the self-correlation
$C(\theta)$; mathematically, this means that some derivative is not continuous at this point.
For our case, $d=2$, this happens in any TPCF such that
\begin{equation}
C(\theta )=\left \{ 
\begin{array}{ll}
g_1(\theta ),& \mbox{$\theta \le \theta_0$}
 \\
g_2(\theta ),& \mbox{$\theta > \theta_0$}
\end{array}
\right \} \;, 
\end{equation}
with $g_1^{(n)}(\theta_0^-)\ne g_2^{(n)}(\theta_0^+)$. 

The exact structure of the oscillations depends both
on the point with some non-continuous derivative, $\theta _0$, and also
on the slope before and after that point: $g_1(\theta_0^-)$ and $g_2(\theta_0^+)$. 
In general, it may also happen that there is more than one point with a point of this
kind. In the case of CMBR analysis, one point with some non-continuous derivative 
is enough to explain all the peaks\cite{Lop12,Bas02,Her04}.

\section{Examples and toy models} 
\label{.examples}

In order to make explicit the discussion in Sect. \ref{.oscill} 
let us consider the following two TPCF $C_1(\theta)$ and $C_2(\theta)$: 
\[
C_1(\theta)=A_{1,1} \exp(-\theta/\theta_{1,1} )+
A_{1,2} \exp(-\theta/\theta_{1,2} )\,,
\] 
where we have chosen the following numerical values:
$A_{1,1}=9744, A_{1,2}=3000, \theta_{1,1}=0.45, \theta_{1,2}=13.0$ (angles are
in degrees), and
\[
C_2(\theta)=\left \{ 
\begin{array}{ll}
A_{2,1}\exp(-\theta/\theta_{2,1} ),& \mbox{$\theta \le \theta_*$}\\
A_{2,2}\exp(-\theta/\theta_{2,2} ),& \mbox{$\theta > \theta_* $}
\end{array}
\right. 
\;, 
\]
where 
$A_{2,1}=12000, A_{2,2}=3600, \theta_{2,1}=0.79, \theta_{1,2}=11.45, \theta_*=1.03$. 

\begin{figure}
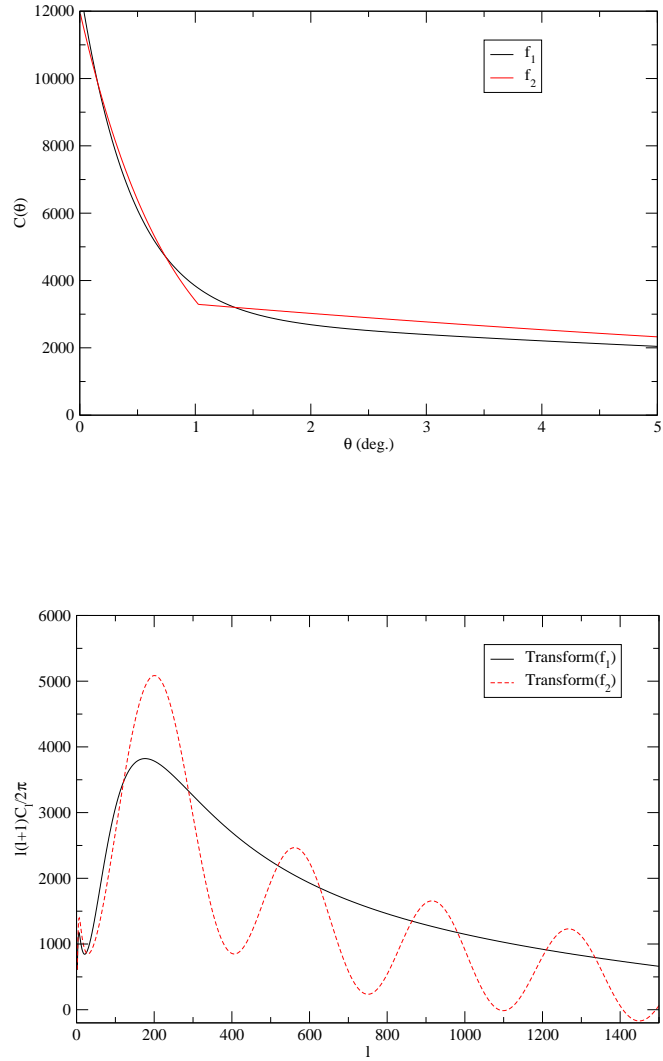

\begin{center}
\vspace{1cm}
\mbox{\epsfig{file=corrtest.eps,height=6cm}}\\
\vspace{2cm}
\mbox{\epsfig{file=power_sp_test.eps,height=6cm}}
\end{center}
\caption{Upper panel: 
Test correlation functions $C_1(\theta)$ and $C_2(\theta)$. Bottom panel: 
The corresponding power spectra $FT[C_1]$ and $FT[C_2]$.}
\label{Fig:corrf1f2}
\end{figure}

In Fig. \ref{Fig:corrf1f2}, we plot these two correlation functions and their corresponding 
power spectra, derived through Eq. (\ref{cl}).

While both correlation functions present a relatively fast change 
in behavior at  $\theta \approx1^\circ $, only the power spectrum 
of $C_2(\theta)$ presents an oscillating series of peaks. 

This simple example clearly shows that only 
the Fourier transform of a function which has
a discontinuous derivative (i.e., $C_2(\theta)$) presents 
a series of oscillating peaks.

Let us now consider two illustrative
  toy models, constructed following the recipe discussed in
  Sect. \ref{.mat1}.

\subsection{Toy model 1: disks with same radius}

We construct a statistically isotropic temperature field $T(\hat\bn)$ as follows. We distribute 
disks of angular radius $R$ such that for separation angles from the center of each disc 
$\theta \le R$ the temperature field is $T(\hat\bn)=f(\theta _r)$, $\theta _r$ being the distance from the 
center of the disk and  the value of the field outside the disks being $T=0$. There are $N_c$ disks in 
a given angular region; their centers are distributed as a spatially homogeneous stochastic point 
process with an angular two-point correlation function $\omega (\theta )$. When $\omega (\theta )=0$ 
for $\theta>0$ the  centers of the disks are distributed as a homogeneous spatial Poisson process. Given that we are
interested in low values of $\theta \ll 1 $ (radians), we can use
 the approximation $\sin (\theta) \approx \theta $.

\begin{figure}
\begin{center}
\vspace{1cm}
\mbox{\epsfig{file=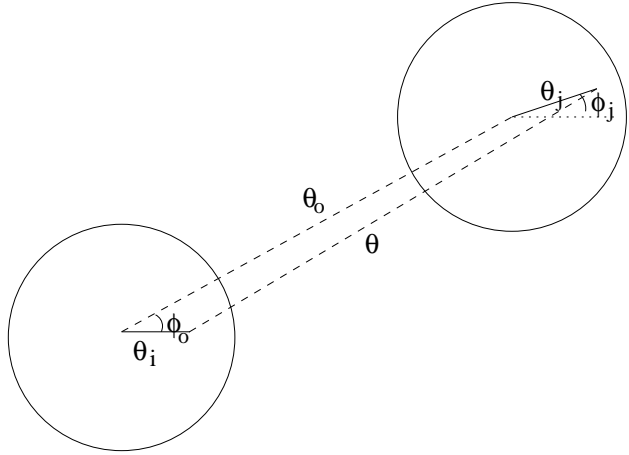,height=6cm}}
\end{center}
\caption{Representation of the geometrical relationships between two
  points $i$, $j$ at two disks.}
\label{Fig:2circles}
\end{figure}

The self-correlation of the field $T$ corresponds to the
 integration over all possible pairs of points $(i,j)$ 
respectively within two disks (see Fig. \ref{Fig:2circles}). The self correlation is the 
average of the product of the value of the temperature
in the first disk, $f(\theta _i)$, multiplied by the value of the temperature in
the second disk, $f(\theta _j)$. The cases in which some point $i$ or $j$ is not within
some disk give a null contribution to the integral. Therefore, our problem reduces to
constraining the limits of the integral of the average correlation within the cases
with non-null contribution. The problem is not trivial but, after a patient analysis,
one can see that the solution is given by:
\bea
&&
C(\theta ) = \frac{N_c}{4\pi\ 2\pi\theta }\int_0^{2\pi}d\phi _i
\\ \nonumber 
&&
\times \int_0^Rd\theta _i\theta_if(\theta
_i)[I_s(\theta, \theta _i)+I_o(\theta,\theta _i)]
,\eea 
where $I_sd\theta $ and $I_od\theta $ stand for the
integrals of the value of $T$ over all pixels $\vec{x}_j$ such that 
$\theta<|\vec{x}_i-\vec{x}_j|<\theta +d\theta$ respectively in the
areas of the same circle of $\vec{x}_i$ ($I_s$) and in the areas of
other  disks ($I_o$). 
They do not depend on $\phi _i$ because both
$f$ and $\omega $ depend only on the angular distance.

$I_s$ is an integral of the second areas in the case that $i$ and $j$ are in the same disk, 
integrating over all possible values of $\theta_j$, $\phi_j$ that keep the distance between the
points $i$ and $j$ between $\theta $ and $\theta+d\theta $. It is:
\bea
&&
\label{Is}
I_s(\theta ,\theta _i) d\theta =\int _0^{2\pi }d\phi _j
\int _{0;\theta<|\vec{x}_i-\vec{x}_j|<\theta +d\theta}^R
d\theta _j\theta _jf(\theta _j)
%
\\ \nonumber 
&& 
=\int _0^{2\pi }d\phi _j \theta _j (\theta, \theta _i,\phi _j)f[\theta
  _j(\theta, \theta _i,\phi _j)]\left| \frac{\partial \theta
  _j(\theta, \theta _i,\phi _j)}{\partial \theta }\right|d\theta . 
\eea
The evaluation of $\theta _j$ is a function of $\theta
  $, $\theta _i$, $\phi _j$, such that $|\vec{x}_i-\vec{x}_j|=\theta
  $:

\bea
&&
\theta _j (\theta, \theta _i,\phi _j)=\theta _i \cos(\phi _j)\pm
\sqrt{\Delta}
\\ \nonumber 
&& \Delta=\theta _i^2\cos(\phi
_j)^2-\theta _i^2+\theta^2
\\ \nonumber  
&&
\mbox{where} \;\; \Delta \ge 0; 0\le \theta _j\le R \;.
\eea
If, within the constraint, there are two
    possible values of $\theta _j$, both will contribute to
    the integral in Eq. (\ref{Is}), while if there are no values of
    $\theta _j$ no value will be included in the integral.  This is a
    special case of Eq. (\ref{thetaj_2c}) for $\theta _o=0$. 

For the evaluation of the second integral involving $I_o$, we have to do something
similar but moving the center of the disk $j$ among all
possibilities. The center of the ``other'' disk is at angular
distance $\theta _o$ from the first one, and azimuth $\phi _o$. Thus,

\begin{equation}
I_o(\theta ,\theta _i )d\theta =\int _0^{2\pi }d\phi _o
\int _0^{\infty }d\theta _o \theta _oP(\theta _o) 
\label{Io}
\end{equation}\[ 
 \times \int _0^{2\pi }d\phi _j
\int _{0;\theta<|\vec{x}_i-\vec{x}_j|<\theta +d\theta}^R d\theta
_j\theta _jf(\theta _j)=
\]\[
\int _0^{2\pi }d\phi _o\int _0^{\infty }d\theta _o \theta _oP(\theta _o)
\]\[\times 
\int _0^{2\pi }d\phi _j \theta _j (\theta, \theta _i,\phi _j,\theta
_o,\phi _o)f[\theta _j(\theta, \theta _i,\phi _j,\theta _o,\phi
  _o)]
\]\[
 \times \left| \frac{\partial \theta _j(\theta, \theta _i,\phi _j,\theta
  _o,\phi_o)}{\partial \theta }\right| d\theta 
\] 
where $P(\theta
_o)$ is the probability per unit area of finding a circle at distance
$\theta _o$ from the center of the first circle. $\theta _j$ is a
function $\theta $, $\theta _i$, $\phi _j$, $\theta _0$, $\phi _o $ standing
again for the value of $\theta _j$, which follows
$|\vec{x}_i-\vec{x}_j|=\theta $ but taking into account that $\theta
_j$ and $\phi _j$ are coordinates with respect to a second circle
whose center is at coordinates $\theta _o$, $\phi _o$ with respect to
the first one. In addition, by definition we have: 
\begin{equation}
P(\theta _o)=\frac{N_c}{4\pi }[1+\omega (\theta _o)] \;. 
\end{equation}

For the calculation of $\theta _j$, by considering the two-dimensional 
planar approximation (instead of spherical geometry) illustrated in
 Fig. \ref{Fig:2circles} we find
\bea
&&
\theta =|\vec{x}_i-\vec{x}_j| \;, 
\\ \nonumber &&
\vec{x}_i=\theta _i\vec{i} \;, 
\\ \nonumber && 
\vec{x}_j=\theta _o(\cos \phi _o\vec{i}+\sin \phi _o\vec{j})
+\theta _j(\cos \phi _j\vec{i}+\sin \phi _j\vec{j}) \;. 
\eea
Hence,
\bea
&&
\theta _j (\theta, \theta _i,\phi _j,\theta _o,\phi _o)=-\theta
_o\cos(\phi _o-\phi _j)+\theta _i \cos(\phi _j)
\label{thetaj_2c}
\\ \nonumber && 
\pm \sqrt{\Delta}
\\ \nonumber && 
\Delta=[\theta _o\cos (\phi _o-\phi _j)-\theta
  _i\cos(\phi _j)]^2-\theta _o^2-\theta _i^2+\theta^2
\\ \nonumber && 
+2\theta _i\theta _o\cos (\phi _o)
\\ \nonumber && 
\mbox{where}\;\;\;     
 \Delta \ge 0; 0\le \theta _j\le R \;.
\eea 
As in the previous case, if there
are two possible values of $\theta _j$, both of them contributing
to the integral in Eq. (\ref{Io}), while if  there are no values of $\theta
_j$ within the constraint, no value will be included in the integral.

\begin{figure}
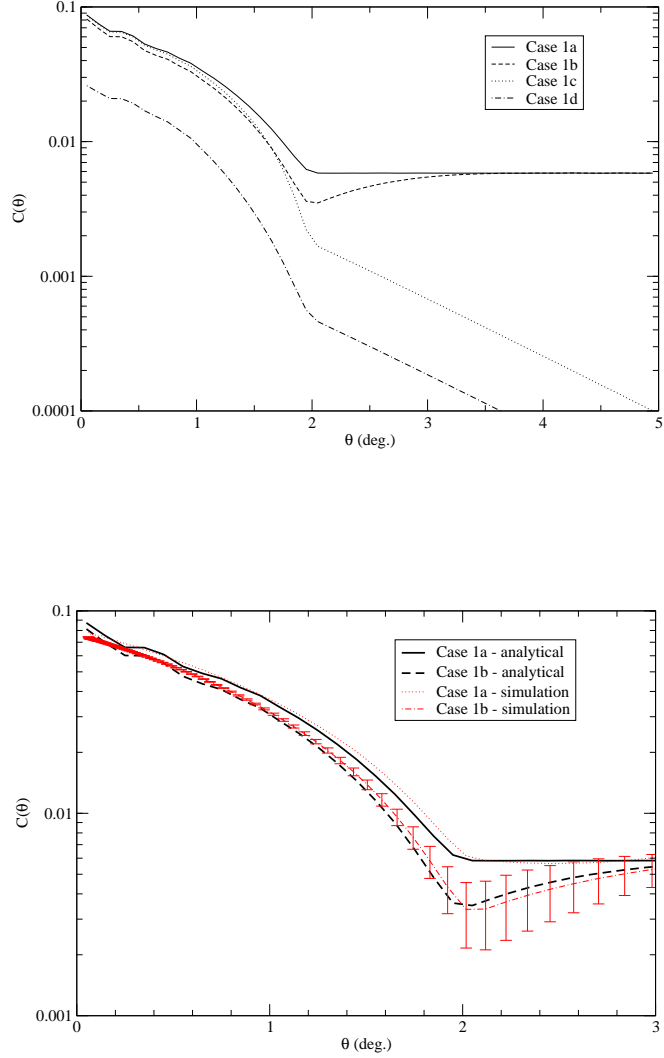

\begin{center}
\vspace{1cm}
\mbox{\epsfig{file=restoy1.eps,height=6cm}}\\
\vspace{2cm}
\mbox{\epsfig{file=restoy1sim.eps,height=6cm}}
\end{center}
\caption{Upper panel: Self-correlation $C(\theta )$ for 
  ``toy-model 1'' cases: a) $f(\theta )=1$, Poissonian distribution of
   disks; b) as in case $a$ but with the added restriction that disks do
  not intersect); c) as in case $a$ with two-point angular correlation
  function of the centers of disks $\omega (\theta )=2\times exp(-\theta
  /R)-1$; d) as in case $c$ but with $f(\theta )=exp(-\theta /R)$. All
  the cases with a number of disks $N_c=10^3$ of radius $R=1^\circ
  $. Bottom panel: Comparison of cases 1a) and 1b) derived
  analytically with the results of Monte Carlo simulations; 
with the error bars, we give the r.m.s. of the simulation 1b); 
for case 1a) is similar.}
\label{Fig:restoy1}
\end{figure}

The results of this toy model for some test functions are given in
Fig. \ref{Fig:restoy1} respectively for: 

a) $f(\theta )=1$ (homogeneous distribution within the disk),
$\omega (\theta )=0$ (Poissonian distribution of
  disks), $R=1^\circ $, $N_c=10^3$;

b)
$f(\theta )=1$, and
\bea
\omega (\theta )= \left \{
\begin{array}{ll}
-1,& \mbox{$\theta\le 2R$}
 \\
0,& \mbox{$\theta > 2R$}
\end{array}
\right \}
\eea 
i.e., a Poissonian distribution, with the added restriction that
 disks do not intersect, $R=1^\circ $, $N_c=10^3$;

c) $f(\theta )=1$, $\omega
(\theta )=2\times \exp(-\theta /R)-1$, $R=1^\circ $, $N_c=10^3$; 

d) $f(\theta
)=\exp(-\theta /R)$, $\omega (\theta )=2\exp(-\theta /R)-1$, $R=1^\circ
$, $N_c=10^3$.  

We have also carried out Monte Carlo simulations: We have randomly
distributed $N_c$ centers, with or without the constraint that minimal distance between two nearest
neighboring centers is larger than $2R$, where $R$ is the disk's size. Inside each disk of size $R$ 
we randomly distribute $N_p$ points. When disks are in a non-overlapping condition the minimal distance 
between the $N_c$ centers has to be $>2R$. We have chosen $R=1^\circ $ and we distribute $N_c=80$ 
disks in a square of side $L=1$ rad., in addition we fixed $N_c=1256$; the
correlation functions are computed as an average over 50 realizations of the distribution.
The results of the analytical calculation of cases 1a) and 1b) were
compared with these Monte Carlo simulations (with the appropriate normalization to get the same
value of $f(\theta <1^\circ )=1$). The results are plotted in Fig. \ref{Fig:restoy1}:
numerical simulations nicely agree with analytical calculations.

\subsection{Toy model 2: disks with varying radii}

With regard to disks of different size, we 
consider only the case in which $\omega(\theta )=0$, i.e.,
a Poisson distribution of disk centers.
In the more general case of $\omega(\theta)\ne 0$, 
 the shape of $C(\theta )$ would be changed, 
as we have seen in the previous subsection. However the
feature corresponding to the abrupt transition [i.e., a 
discontinuous first derivative of $C(\theta )$], which is the scale of interest for the 
generation of the oscillating sequence of peaks 
in the angular power spectrum, does not depend on the particular shape of $\omega (\theta )$. 
We are going to show that the angular TPCF of the set of disks of different size  
is again characterized by a discontinuous derivative.

Let us assume that we have disks of different angular sizes $R$, and that, 
conditioned by their size, the contribution to the angular correlation function of disks with radius
 between $R$ and $R+dR$ follows a dependence
\begin{equation}
C(\theta ,R)=
\left \{ 
\begin{array}{ll}
A(R)h(\theta /R),& \mbox{$\theta\le 2R$}
 \\
0,& \mbox{$\theta > 2R$}
\end{array}
\right. \; ,
\end{equation} 
where $A(R)$ is the amplitude
of the correlation of disks with radius $R$, and
  $h$ is a smooth continuous function with all continuous derivatives
  such that $h(2)=0$; such a distribution of $h$ is given by the contribution to the correlation
of the disks with the same size (i.e., the toy model discussed in the previous
  section); we subtract the average field [thus, the constant $C(\theta >2R)$ may
  be set as zero conveniently defining a new $T^*=T-constant$].
Assuming a uniform distribution of radii between $R_{\rm min}$ and $R_{\rm max}$,
given that there is no correlation between the positions of the
  disks, the angular TPCF is just the sum of the
correlation function between its minimum and maximum
radii. 
That is,
\bea
&&
C(\theta )=\int _{R_{min}}^{R_{max}}C(\theta, R)dR
=
\\ \nonumber 
&&
\left \{ 
\begin{array}{ll}
\theta \int _{\theta /R_{max}}^{\theta /R_{min}} dx A(\theta /x)h(x)/x^2,& \mbox{$\theta\le 2R_{min}$}
 \\
\theta \int _{\theta /R_{max}}^{2} dx A(\theta /x)h(x)/x^2,& \mbox{$2R_{min}<\theta < 2R_{max}$}
 \\
0,& \mbox{$\theta \ge 2R_{max}$}
\end{array}
\right. \;. 
\eea 
If there are 
  minimum and maximum radii there is clearly a change of dependence
  with $\theta $ for three regimes. If, instead, there is only a
  minimum ($R_{min}>0$) or a maximum radius, there will be two
  regimes. Finally, if $R_{min}=0$, $R_{max}=\infty $, then there will be
  only one regime and no abrupt transition.

Let us consider an example: a) $A=1$,
  $h(x)=(1-x/2)$, for which we get
\begin{equation}
C(\theta )=
\label{ccc} 
\end{equation}\[
\left \{ 
\begin{array}{ll}
(R_{max}-R_{min})-\frac{1}{2}ln(R_{max}/R_{min})\theta ,& 
\mbox{$\theta\le 2R_{min}$}
 \\
R_{max}-\frac{1+ln(2)}{2}\theta+\frac{1}{2} 
ln(\theta /R_{max}),& \mbox{$2R_{min}<\theta < 2R_{max}$}
 \\
0,& \mbox{$\theta \ge 2R_{max}$}
\end{array}
\right. 
\;.
\]

\begin{figure}
\begin{center}
\vspace{1cm} 
\mbox{\epsfig{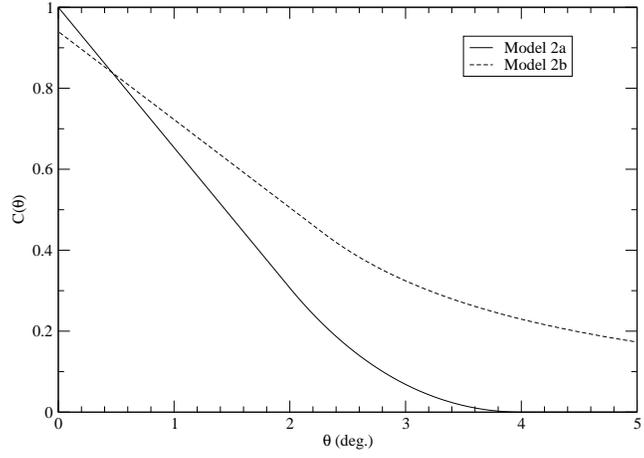}}\\
\vspace{2cm}
\mbox{\epsfig{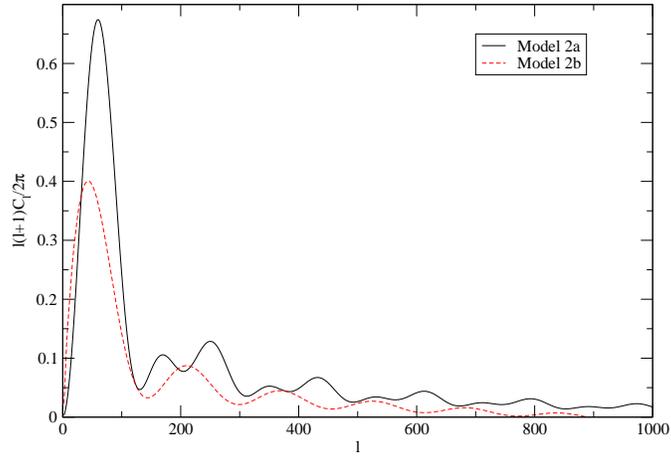}}
\end{center}
\caption{Upper panel: Two-point correlation $C(\theta )$
  for the disks of varying radius, when no correlation among
    disk centers is considered a), b). Bottom panel: The
  corresponding power spectra.}
\label{Fig:restoy2}
\end{figure}

Fig. \ref{Fig:restoy2} displays Eq. (\ref{ccc}) for $R_{min}=1^\circ$
and $R_{max}=2^\circ$.

To each disk radius projected on the sky we could associate the
  physical distance $R=L/r$, where $r$ is the distance and $L$ is
  the physical size in an Euclidean space. As it will be discussed in \cite{Lop12},
this may be representative of spheres placed at different distances.
For $A(R)=\left(\frac{A_0}{R}\right)^2$ and 
for instance $h(x)=(1-x/2)$, we find 

\begin{equation}
\label{cccc} 
C(\theta )=
\end{equation}\[
\left \{ 
\begin{array}{ll}
A_0^2[\frac{r_{max}-r_{min}}{L}-\frac{1}{4L^2}(r_{max}^2-r_{min}^2)\theta ] ,
& \mbox{$\theta \le \theta _1$}
 \\
A_0^2[-\frac{r_{min}}{L}+\frac{1}{\theta }+\frac{r_{min}^2}{4L^2}\theta ],
& \mbox{$\theta _1<\theta < \theta _2$}
 \\
0,& \mbox{$\theta \ge \theta _2$}
\end{array}
\right.
,\]\[
\theta _1\equiv 2L/r_{\rm max}; \ \ \theta _2\equiv 2L/r_{\rm min}
\]

Fig. \ref{Fig:restoy2} displays 
Eq. (\ref{cccc}) for a case b) $A_0=0.02$, $r_{min}/L=3$,
$r_{max}/L=50$.

As  can be observed, the correlation function is smoother.
However, the power spectra (Fig. \ref{Fig:restoy2}) 
still show a sequence of oscillating peaks.
Indeed, another way to understand the fact that
oscillations are still present when the disk size is not fixed is 
obtained recalling that the power spectrum is
the superposition of the power spectra for each radius (see \S \ref{.mat2}): 
the superposition of oscillating power spectra
gives still a global oscillation, provided they do not cancel
each other out (which is the case when there is no minimum and/or
maximum radius).

The examples of this subsection are some particular ways of convolution. In general,
any kind of convolution or smoothing in the abruptness of the disc shape will also
produce oscillations, though with damped amplitude of the oscillation.

\section{Conclusions}
\label{.concl}

Although the power spectrum analysis is useful for obtaining some kind of information on CMBR anisotropy data,
the angular TPCF (Two-point correlation function) analysis gives a more straightforward physical 
representation of that distribution. Moreover, while the power spectrum shows repeated information
 in the form of its multiple peaks and oscillations, the TPCF, its Fourier transform, offers a 
 compacter presentation which condenses all the information of the multiple peaks in a localized 
 real space feature. 
Precisely because of that, and because there is a dearth of literature analyzing the properties of 
the TPCF, we have concentrated here on its analysis.

In this paper we have shown how oscillations in the power spectrum arise when there are some point 
singularities in the angular TPCF  in the form of some discontinuity point of its $n$-th derivative 
at some angular distance. In particular, we have clarified the analytical link between 
the frequency and the power law order decay of damped oscillations in the power spectrum and the kind 
of singularity in the TPCF.
We have then presented toy models able to generate and illustrate this phenomenology: 
simply by placing some distribution in the sky of filled disks of fixed or variable radii with an 
excess of antenna temperature, provided there is a minimum$>0$ and/or a maximum in the range of 
radii.  Disks of different radii may stand for projection of spheres with variable radii or spheres 
with variable/fixed radii and variable distance from the observer. 

Discussions on the physical interpretation of these mathematical properties for the
case of CMB anisotropies and comparison with real CMBR data will be given in \cite{Lop12}. 
This paper, now in preparation, will discuss these mathematical properties 
in terms of matter distribution in the fluid which is generating the radiation. 
The angular correlation function of the anisotropies from CMBR data will also
be calculated, in an effort to derive the minimum number of parameters of a generic function to fit it. 
The paper \cite{Lop12} will argue that a power spectrum with oscillations is a rather normal characteristic 
expected from any fluid with clouds of overdensities that emit/absorb radiation 
or interact gravitationally with the photons, and with a finite range of sizes 
and distances for those clouds. It will also show 
that the angular correlation function can be fitted by a generic function 
with a total of $\approx 6$ free parameters. The standard cosmological interpretation of ``acoustic'' peaks
is just a particular case; peaks in the power spectrum might be generated in 
scenarios that have nothing to do with oscillations due to gravitational compression 
in a fluid. Nonetheless, the standard model with six free parameters produces a better fit than 
the generic fit with the same number of free parameters; it fits third and higher order peaks whereas 
the generic fit reproduces only the first two peaks.

Apart from the analysis of CMBR peaks, the concepts analyzed in this paper are general and applicable 
to any kind of power spectrum with oscillations/peaks. The BAO oscillations in the large-scale 
structure\cite{Kaz10} might be, for instance, another field where these results can be applied.

\section*{Acknowledgments}
Thanks are given to: F. Sylos Labini, for providing the results of the Monte Carlo simulation
given in Fig. 3/bottom, and for his many suggestions on the text of this paper;
F. Atrio-Barandela and the anonymous referee, for their helpful comments;
T. J. Mahoney, for proof-reading this paper. MLC was supported by the grant
AYA2007-67625-CO2-01 of the Spanish Science Ministry.


\end{document}